\documentclass[preprintnumbers, pra, showpacs, floatfix,twocolumn,
preprintnumbers, letterpaper, superscriptaddress]{revtex4}
\usepackage{amsfonts}
\usepackage{amsmath}
\usepackage{amssymb,epsf}
\usepackage{latexsym}
\usepackage{graphicx,epsfig}
\usepackage{amssymb}
\usepackage{subfigure}
\usepackage[colorlinks=true,citecolor=blue,linkcolor=blue,urlcolor=black]{hyperref}
\usepackage{epstopdf}
\usepackage{color}

\def\be{\begin{equation}}
\def\ee{\end{equation}}
\def\ba{\begin{eqnarray}}
\def\ea{\end{eqnarray}}
\def\la{\langle}
\def\ra{\rangle}

\begin{document}
\title{Partial Anyon Condensation in the Color Code: A Hamiltonian Approach}
\author{Mohsen Rahmani Haghighi}
\email{rahmani.qit@gmail.com}
\affiliation{Physics Department, College of Sciences, Shiraz University, Shiraz 71454, Iran}
\author{Mohammad Hossein Zarei}
\email{mzarei92@shirazu.ac.ir}
\affiliation{Physics Department, College of Sciences, Shiraz University, Shiraz 71454, Iran}
\begin{abstract}
	Lattice Hamiltonians, which can be tuned between different topological phases, are known as important tools for understanding physical mechanism behind topological phase transitions. In this paper, we introduce a perturbed Color Code Hamiltonian with a rich phase structure which can be well matched to the mechanism of anyon condensation in the Color Code. We consider Color Code model defined on a three-colorable hexagonal lattice and add Ising interactions between spins corresponding to edges of the lattice. We show that Ising interactions play the role of physical factor for condensing anyons in the Color Code. In particular, corresponding to three different colors of edges in the hexagonal lattice, we consider three different coupling parameters. Then, we are able to condense anyons with different colors by tuning power of Ising interactions in the corresponding edges. In particular, we explicitly show that condensation of one type of anyons in the Color Code leads to a phase transition to the Toric Code state. On the other hand, by condensing two types of anyons, we observe a phase transition to a modified version of the Toric Code where partial set of anyons in the Toric Code are condensed and we call it a partially topological phase. Our main method for derivation of the above results is based on a suitable basis transformation on the  main Hamiltonian in the sense that our model is mapped onto three decoupled transverse-field Ising models, corresponding to the three colors. We use the above mapping to analyze behavior of string order parameters as non-local indicators of topological order. We introduce three string order parameters that can well characterize different phases of the model. Specifically we give a simple description of the partially condensed phase by using string order parameters.

\end{abstract}
\maketitle
\section{Introduction}
Topological quantum systems have attracted much attention because of their unique properties including non-local order in the ground state, exotic statistics in point- like excitations and applications in fault-tolerant quantum computation\cite{Xiao1,Wen1,Wen2,Wen5,Kitaev,Nielsen,Nayak}. Because of their non-local nature, topological phases go beyond the Landau theory\cite{Haegeman}. In particular, topological orders can not be classified by symmetric groups as symmetry breaking phases are described. Topological phases are described by unitary modular tensor categories, which can capture exotic properties such as fusion and braiding of excitations \cite{t1, t2}.

Among different topological systems, topological quantum codes \cite{Fowler,Preskill,Freedman, Levin} which are tools for protecting quantum information against local errors\cite{Bravyi,Jamadagni}, are known also as powerful models for exploring topological phases of matter \cite{Gottesman, Brown}. Color Code\cite{Bombin1,Kubica,Ohzeki} and Toric Code \cite{Kitaev} are prominent examples of exactly solvable topological models that exhibit distinct patterns of long-range entanglement \cite{kargar,Kitaev2,Zarei1}. It has been shown that phase transition out of these simple topological phases \cite{phase0,phase1,phase2,phase3,phase30,phase4} and even the transition from the Color Code to the Toric Code \cite{phase6} offers insights into the nature of topological order, long-range entanglement, and the underlying anyonic properties.

Topological quantum codes play also important role in physical understanding of topological phases. Topological order in the Toric Code state is a simple example of a string-net condensation \cite{Levin} which is a physical mechanism for topological phases. The ground state of the Toric Code is a loop-condensed state and excitations including charge and flux anyons correspond to end-points of open strings \cite{exp}. On the other hand, a transition from Toric Code state to a trivial phase can be described by condensation of anyons where the ground state would be a superposition of loops as well as open strings \cite{con}. It is shown that anyon condensation\cite{Bais,Burnell,Kong,duiven,Iqbal,Bombin2} is a powerful framework for describing topological phase transitions where condensing a subset of anyons leads to the confinement of other anyons. It effectively reduces the number of independent anyonic excitations in the system. For example, it is shown that condensation of one type of anyons in the Color Code state leads to a phase transition to the Toric Code state \cite{Kesselring}.

In order to better understanding the mechanism of anyon condensation, it is important to use an approach based on lattice Hamiltonians \cite{lattice,lattice1,lattice2,lattice3} in the sense that phase transitions occur by tuning parameters in the Hamiltonian. In this paper, we consider such approach for hexagonal Color Code model which has recently studied based on anyon condensation \cite{Kesselring}. we introduce a Hamiltonian that consists of the standard Color Code model with additional Ising interaction terms. We control power of Ising terms by tuning coupling parameters $J_c$ that $c=r , b, g$ refer to the red, blue, and green edges of the Color Code, respectively. We show that, by varying $J_c$'s corresponding to different colors from 0 to 1, the system undergoes transitions from the Color Code to different interesting phases. We specifically interpret the above transitions by the mechanism of anyon condensation in the sense that Ising interactions corresponding to each edge of the lattice are responsible for condensing corresponding anyons. We then characterize different phases out of the Color Code in the model. We show there are Toric Code phases, a trivial phase and intermediate phases in which the system shows a partially topological phase corresponding to a modified version of the Toric Code with holes. In this regard, our analysis reveals that by tuning the $J_c$ parameters independently for each color, we effectively implement partial anyon condensation which allows for the emergence of intermediate phases with partially topological order. 

To derive the above results, we use suitable basis transformations on Initial Hamiltonians for different values of $J_c$. In particular, we show that the main Hamiltonian is mapped onto three decoupled transverse-field triangular Ising models with well-known phase transitions \cite{Pfeuty,Blote,Hamer,Croo}. We also use other suitable basis transformations in order to characterize the Toric Code phase as well as the partially topological phase in the model. Finally, we employ non-local string order parameters to capture the topological nature of different phases of the model. To this end, we define three types of string operators corresponding to three colors in the Color Code. Then, using a mapping to Ising order parameters we analyze behavior of string order parameters along phase transition lines. We show that all phases of the system are well characterized by the above three string order parameters.

structure of this paper is as follows. In Sec. \ref{sec1}, we start with a introduction to the Color Code and Toric Code. We then give a brief explanation on anyon condensation in the above models. In Sec. \ref{sec2}, we present our main Hamiltonian and show how it can be mapped to the three triangular Ising models in transverse field. In Sec. \ref{sec20}, we explore the phase diagram of the perturbed Color Code by studying a specific limiting case where Toric Code phase emerges. We specifically explain how Ising perturbations in our Hamiltonian lead to anyon condensation in the Color Code. In Sec. \ref{sec3}, we characterize a partially topological phase in the phase space. In Sec. \ref{sec4}, we introduce the string order parameters for detecting topological nature of different phases of the model.

\section{Topological Quantum Codes And Anyon Condensation:}\label{sec1}

In this section, we give a brief review on the Color Code and Toric Code and then explain how anyon condensation describes topological phase transition out of these topological quantum codes.

\begin{figure}[h!]
\centering
\includegraphics[width=8.5cm,height=4cm,angle=0]{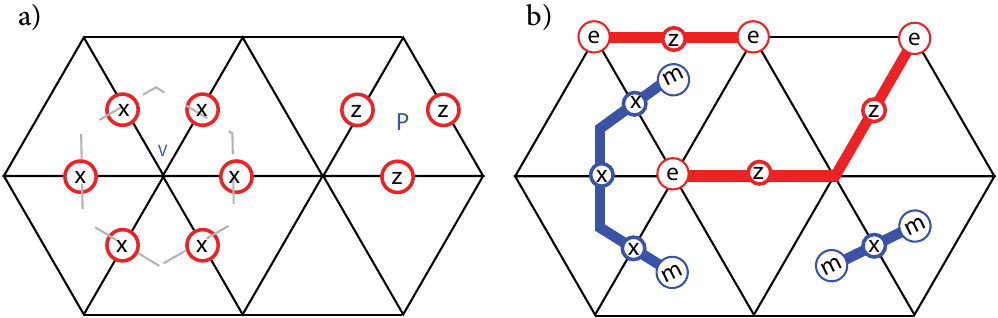}
\caption{(Color online) a) Vertex and plaquette stabilizers in a triangular Toric Code. Vertex operators can be represented by a loop on dual hexagonal lattice. b) Excitations in the Toric Code are generated by applying string of $X$ or $Z$ operators. Flux and charge anyons denoted by $m$ and $e$ live in endpoints of $X$-type and $Z$-type strings, respectively. } \label{A1}
\end{figure} 

Toric Code is a well-known topological quantum code defined on an arbitrary lattice attached on a torus, with qubits placed on edges. The Hamiltonian consists of vertex and plaquette operators:
\begin{equation}\label{HTC}	
H_{TC}=-\sum_v A_v -\sum_p B_p ,
\end{equation}
where $A_v$ and $B_p$ are given by,
\begin{equation}\label{STAB}
B_p =\prod_{i\in \partial p} Z_i ~~,~~A_v =\prod_{i\in v}X_i ,
\end{equation}
where $X$ and $Z$ are Pauli operators and $i\in \ v$ refers qubits located on edges incoming to a vertex and $i\in \partial p$ refers to qubits living on edges around a plaquette. For example, in Fig. \ref{A1}-a, we show these stabilizers for a triangular lattice. The ground state of the Toric Code has a four-fold degeneracy \cite{Kitaev}. In particular, one of the ground states can be expressed by vertex operators in the following form up to a normalization factor:
\begin{equation}\label{GSTC}
    |GS \rangle _{TC}= \prod_{v}(1+A_v) |0\rangle^{\otimes L} ,
\end{equation}

 where $L$ referes to number of qubits. As shown in Fig. \ref{A1}-a, each $A_v$ operator can be represented by a hexagonal loop in dual lattice. Therefore, $\prod_{v}(1+A_v)$ is equal to a superposition of loop operators and accordingly $ |GS \rangle _{TC}$ is described as a condensed state of loops.
 
  Excitations of the Toric Code are called anyons. As shown in Fig. \ref{A1}-b, they are generated by applying a string of $X$ or $Z$ operators in the sense that anyons live in the end-points of strings. There are four types of anyons in the Toric Code including vacuum,  two bosonic anyons (electric charge $e$ and magnetic charge $m$) and their fusion denoted by $f$ ($f=e\times m$), which is a fermionic particle. the fusion and the braiding between anyons are given by the following relations:
\begin{equation}\label{anyontc}
e \times e = 1, \quad m \times m = 1, \quad e \times m = f
\end{equation}
\begin{equation}\label{Braidtc}
M_{u,u} = 1, \quad M_{u,v} = -1 \quad \text{for } u \neq v, \text{ where } u,v \in \{e, m\}.
\end{equation}
Here, $M$ denotes the Monodromy matrix and $ M_{u,v}$ represents the phase acquired when an anyon of type $u$ is braided around an anyon of type $v$. This implies that the mutual braiding between  $e$ and $m$  is nontrivial, while the self-braiding of all anyons is trivial.

\begin{figure}[h!]
\centering
\includegraphics[width=8cm,height=6cm,angle=0]{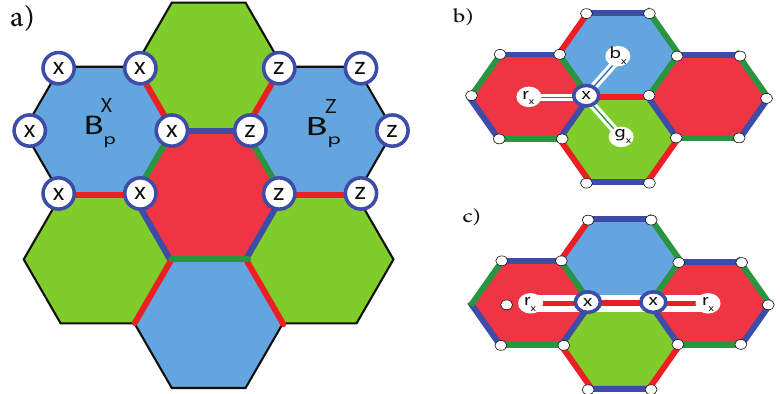}
\caption{(Color online) a) Color Code on a three-colorable hexagonal lattice. $X$-type and $Z$-type Stabilizers are defined corresponding to each plaquette. b) An $X$ operation on a single qubits generates three excitations in three plaquettes with different colors. c) Anyons move in plaquettes with the same color by applying a string of $X$ operators.  } \label{A2}
\end{figure}

On the other hand, the Color Codes are defined on three-colorable lattices with qubits living in vertices. As shown in Fig. \ref{A2}-a for a hexagonal lattice, plaquettes in a three-colorable lattice are assigned by three different colors, and plaquettes with similar color are connected by edges with the same color. Hamiltonian for Color Code is given by:
\begin{equation}\label{HCC}
H_{CC}= -\sum_p B_p^z -\sum_p B_p^x ,
\end{equation}
where $B_p^x$ and $B_p^z$ refere the product of Pauli-X and Pauli-Z operators over all qubits belongings to a plaquette, see Fig. \ref{A2}-a. On a torus, Color Code shows a 16-fold degeneracy in the ground state. Similar to the Toric Code one of the ground states is written in the following form, up to a normalization factor:
\begin{equation}\label{GSCC}
    |GS \rangle _{CC} = \prod_{p}(1+B_p^x) |0\rangle^{\otimes N},
\end{equation}
where $N$ refers to number of qubits. This model also exhibits topological order and supports a rich structure of anyonic excitations. If we represent each $B_p ^x$ operator by a loop similar to the Toric Code, the ground state can be described as a condensed state of loops. However, loops in the Color Code have different colors. The role of color in the Color Code is also seen in the excitations. Specifically, there are 16 anyons, including 9 bosonic, 6 fermionic, and a trivial vacuum state \cite{Kargarian}. The bosonic anyons can be categorized according to their colors and Pauli-type labels in the form of $r_x , r_y , r_z , b_x , b_y , b_z , g_x , g_y , g_z$, where $ r , b , g$ refer to the red, blue, and green plaquettes, respectively, and the subscripts $ x, y , z$ denote the corresponding Pauli operators. As shown in Fig. \ref{A2}-b, anyons are generated by applying strings of $X$ or $Z$ operators. For example, when a single Pauli $X$ ($Z$) error occurs on a qubit , it violates the $Z$-type ($X$-type) stabilizers of the adjacent plaquettes, resulting in three excitations including $r_x , b_x , g_x$ ($r_z , b_z , g_z$)  , as shown in Fig. \ref{A2}-b. On the other hand, when two $X$ ($Z$) operators are applied on two qubits located at the endpoints of an edge, these errors violate the $B^p_z$'s ($B^p_x$'s) associated with the two plaquettes which are adjacent to that edge Fig. \ref{A2}-c. Sequential application of Pauli operators allows the excitations to move throughout the lattice and anyons live in end-points of strings.

Let $u,v,w \in \{r, g, b\}$ refer to colors (red, green, blue), and $ \alpha, \beta, \gamma \in \{x, y, z\} $ refer to the Pauli labels. Then, the fusion rule for the Color Code is given by:

\begin{equation}\label{anyoncc}
u_\alpha \times v_\beta =
\begin{cases}
\mathbf{1}, & \text{if } u = v \text{ and } \alpha = \beta \\
u_\gamma, & \text{if } u = v \text{ and } \alpha \neq \beta \neq \gamma \\
w_\alpha, & \text{if } u \neq v \text{ and } \alpha = \beta \quad \text{and } w \neq u,v \\
w_\gamma, & \text{if } u \neq v \text{ and } \alpha \neq \beta \neq \gamma \text{ and } w \neq u,v
\end{cases}
\end{equation}

For bosons $ u_\alpha $ and $ v_\beta $, the braiding matrix between them is defined in the following form:

\begin{equation}\label{braidcc}
M_{u_\alpha , v_\beta} =
\begin{cases}
+1 & \text{if } u = v \text{ or } \alpha = \beta \quad \text{(trivial braiding)} \\
-1 & \text{otherwise} \quad \text{(non-trivial braiding)}
\end{cases}
\end{equation}

All properties of topological quantum systems reflect in braiding and fusion of anyons. In particular, phase transitions between different topological phases can be described in terms of anyons. An interesting approach in this direction is anyon condensation where bosonic anyons are condensed in the ground state \cite{Kesselring}. As a result, anyons that have nontrivial braiding with the condensed bosons become confined. It means they cannot freely propagate without an energy cost. Anyon condensation leads to a topological phase transition where initial set of anyons is replaced by a new set corresponding to a new topological phase. Here, we explain this idea by using examples of Toric Code and Color Code.

In the Toric Code model, we can consider condensation of flux anyons $m$. It means that in the ground state we have a condensation of loops as well as open strings with flux anyons in their endpoints. Such a state can be simply written in the following form:
\begin{equation}
	|\psi\rangle= \prod_i (1+X_i) |0\rangle ^{L}
\end{equation}
In other words, each term in $\prod_i (1+X_i)$ is a product of $X_i$'s which is equal to a configuration of $X$-type closed or open strings. However, $|\psi \rangle$ is equal to a trivial state of $|+\rangle ^L $ and therefore, condensation of flux anyon leads to a phase transition from a Toric Code phase to a trivial phase. In terms of behavior of anyons, when we condense flux anyons, other anyons are confined because they have a non-trivial braiding with $m$. In this regard, charge anyon $e$ and fermion $f$ should be removed from the initial set of anyons. On the other hand, flux anyon $m$ is also replaced by vacuum because it is now condensed in the sense that the effect of $X$ operator on the ground state is trivial. In this regard, in the new set of anyons,  it remains a vacuum particle and therefore we have a trivial phase.
 
Anyon condensation can also describe a phase transition from Color Code to Toric Code. To this end, it is enough to condense the anyon $r_x$ in the Color Code where the new ground state becomes a superposition of loops with different colors in addition to open strings with red color. As a consequence, anyons that have nontrivial braiding with $r_x$ including $g_y, g_z, b_y,$ and $ b_z$ become confined and they should be removed from the initial set of anyons. On the other hand, we notice that in the fusion rules, the condensed anyon $r_x$ should be replaced by vacuum. For example, in the fusion rule $g_x \cdot r_x = b_x$, $r_x$ is replaced by $1$ and consequently $g_x$ and $b_x$ effectively are the same anyon. A similar argument holds for $ r_y $ and $ r_z $.  Finally, this condensation reduces the number of independent bosonic anyons to two, which matches with the number of bosons in the Toric Code. In this regard, condensation of $r_x$ leads to a transition from Color Code to the Toric Code \cite{Kesselring}. 

Although the above interesting arguments on anyon condensation provide a simple description of topological phase transitions, it is also an important task to support this mechanism by a Hamiltonian approach where the corresponding phase transitions occur by tuning physical parameters in a suitable Hamiltonian. In the following of the paper, we introduce a perturbed Hamiltonian of the Color Code to support the above mechanism of anyon condensation. Furthermore, using the Hamiltonian approach helps us to discover other possible topological phases which can be generated by anyon condensation in the Color Code.  

\section{ Color Code in presence of Ising interactions:}\label{sec2}
We consider a modified Hamiltonian of the hexagonal Color Code where we add Ising interactions between spins corresponding to each edge of the lattice:
\begin{equation}\label{Hamiltonian}
\begin{aligned}
H = & - \sum_{p=r,g,b} B_p^{x} 
- \sum_{c=r,g,b} (1 - J_c) \sum_{p \in c} B_p^z \\
& - \sum_{c=r,g,b} J_c \sum_{\langle i,j \rangle \in c} X_i X_j ,
\end{aligned}
\end{equation}

where $X_i X_j$ refers to an Ising interaction and $\langle i,j \rangle\in c$ refers to neighboring qubits corresponding to an edge with color $c$. $J_c$'s, which take values from 0 to 1, refer to coupling parameters corresponding to edges with three colors $c=r,g,b$ .

\begin{figure*}[htbp]
	\centering
	\includegraphics[width=\textwidth]{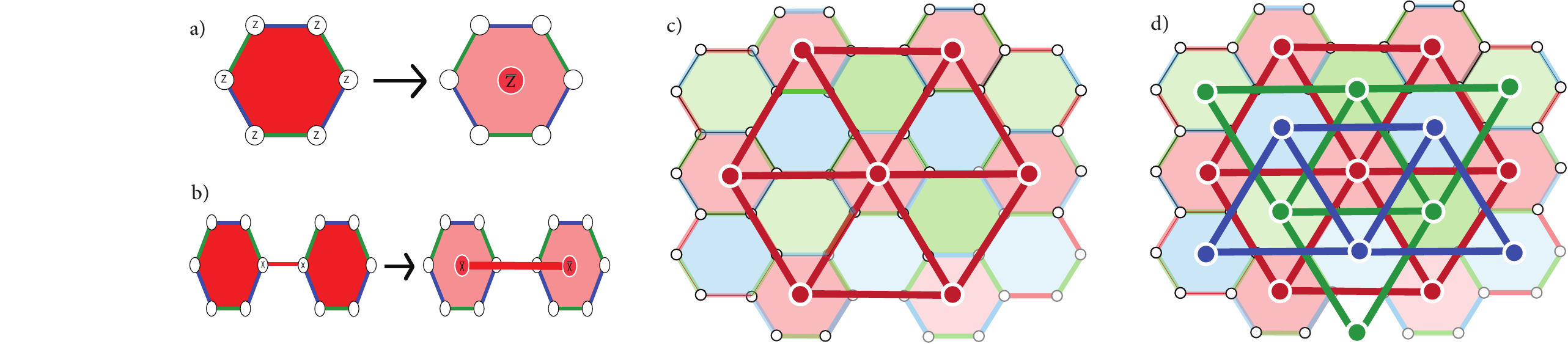}
	\caption{(Color online) a) A $Z$-type stabilizer, which is applied to six physical qubits in the Color Code,  is mapped to a single phase operator applied to an Ising qubit in the center of the corresponding plaquette. b) An Ising interaction between two physical qubits is mapped to a similar interaction between Ising qubits living in the corresponding plaquettes. c) Interaction terms corresponding to the red color in the initial Hamiltonian are mapped to a transverse-field Ising model defined on Ising qubits designed on a triangular lattice. d) Other interaction terms in the initial Hamiltonian are also mapped to  transverse-field Ising models on triangular lattices with different colors. There are finally three decoupled Ising models defined on Ising qubits with three different colors.  }\label{A3}
\end{figure*}

In order to characterize the phase structure of this model, we apply a basis transformation that maps the initial Hamiltonian to a familiar quantum model. To this end, consider a new basis that we call Ising basis, defined as :
\begin{equation}\label{basis1}
 \prod_{p} |\mu_p\ra = \prod_{p} (1+(-1)^{\mu_p} B_p^{z}) |+\ra^{\otimes N} ,
\end{equation}
where we ignore the normalization factor and $\prod_p$ refers to product of all $N/2 -1$ independent plaquette operator $B_p$. each $\mu_p$ takes binary values $0$ and $1$ and therefore, it can be interpreted as a logical qubit living in the center of the plaquette $p$ and we call it an Ising qubit. We denote Ising qubits by red, green and blue circles, as illustrated in Fig. \ref{A3}. Notice that the above basis is not complete because the number of qubits is $N$ which is two times of the number of plaquettes. In fact one should also use stabilizer operators $B_p ^x$ to define other bases for the Hilbert space. However, since Ising interactions commute with $B_p ^x$'s, we only consider a subspace of the Hilbert state which is constructed by the $B_p^z$ operators.

Now, we rewrite the Hamiltonian (\ref{Hamiltonian}) in the Ising basis (\ref{basis1}). Notice that  since the change of basis does not change the spectrum of the initial  Hamiltonian, phase structure of the final Hamiltonian is the same as phase structure of the initial Hamiltonian. In this regard, we consider the effect of different terms in the initial Hamiltonian in the Ising basis. The operator $B_p^{x}$ in Hamiltonian (\ref{Hamiltonian}) commutes with all $B_p^{z}$ operators and therefore for a plaquette $p'\in c$, we have:
\begin{equation}\label{basis2}
 B_{p'}^{x} \prod_{p} |\mu_p\ra = [\prod_{p} (1+(-1)^{\mu_p} B_p^{z}) ] B_{p'}^{x} |+\ra^{\otimes N} .
\end{equation}
Then since $ B_{p'}^{x} |+\ra^{\otimes N}=|+\ra^{\otimes N}$, it is concluded that $ B_{p'}^{x} \prod_{p} |\mu_p\ra = (+1)  \prod_{p} |\mu_p\ra$. It means that the first term in the Hamiltonian (\ref{Hamiltonian}) is equal to a constant energy shift $H_0 = -N$ in the Ising basis (\ref{basis1}).

We then consider the effect of the next term of the Hamiltonian including $B_p^z$ in the Ising basis. For example, for a specific red plaquette denoted by $p\prime$, shown in Fig. \ref{A3}-a, we have:

$$	B_{p'}^z \prod_{p} |\mu_p\rangle
= B_{p'}^z \left(1 + (-1)^{\mu_{p'}} B_{p'}^z \right)$$
\begin{equation}
	\label{B}\quad \times \prod_{p \neq p'} \left(1 + (-1)^{\mu_p} B_p^z \right) |+\rangle^{\otimes N}
\end{equation}

Then since $[B_{p'}^z ]^2 =1$, it is clear that $B_{p'}^z \left(1 + (-1)^{\mu_{p'}} B_{p'}^z \right)=(-1)^{\mu_{p'}} \left(1 + (-1)^{\mu_{p'}} B_{p'}^z \right)$. Accordingly, we conclude that $ B_{p'}^{z} \prod_{p} |\mu_p\ra = (-1)^{\mu_{p'}}  \prod_{p} |\mu_p\ra$. It means that the operator $B_{p'}^z$ is equal to a logical phase operator $\bar{Z}$ acting on the Ising qubit located at the plaquette $p'$, see Fig. \ref{A3}-a.

 Thus, by rewriting other plaquette operators in the same way, the term of
$-\sum_{c=r,g,b} (1 - J_c) \sum_{p \in c} B_p^z $ in the Hamiltonian is transformed to $H_I= -\sum_{c=r,g,b} (1-J_c)\sum_{i\in c} \bar{Z_i}$ in the Ising basis where $i \in c$ refers to an Ising qubit living in plaquettes with color $c$. 

Next, we examine the effect of the Ising terms in the initial Hamiltonian (\ref{Hamiltonian}) on the Ising basis. First, let us focus on the Ising interaction corresponding to a red edge, as shown in Fig. \ref{A3}-b. Notice that the operator $X_i X_{i+1}$, which is applied to two qubits located at the end-points of a red edge, anti-commutes with the $B_{p'}^z$ and $B_{p'+1}^z$ corresponding to two plaquettes which are connected by a red edge, see Fig. \ref{A3}-b.  Therefore, since $X_i X_{i+1}B_{p'}^z =(-1) B_{p'}^z X_i X_{i+1}$ and by the fact that $X|+\ra = (+1)|+\ra$ it is concluded that:
$$X_i X_{i+1} \prod_p |\mu_p\rangle =$$
$$\left(1 + (-1)^{\mu_{p'} + 1} B_{p'}^z )(1 + (-1)^{\mu_{(p'+1)} + 1} B_{p'+1}^z \right)\times\nonumber \\
  \prod_{p \notin \{p', p'+1\}} $$
\begin{equation}
\left(1 + (-1)^{\mu_p} B_p^z \right) |+\rangle^{\otimes N}
\end{equation}

 It means that the effect of $X_i X_{i+1}$ in the Ising basis is equal to two logical shift operators $\bar{ X}$ which shift red Ising qubits $\mu_{p'}$ and $\mu_{p'+1}$ to ${\mu_{p'}} + 1$ and $\mu_{(p'+1)}+1$ respectively. In this regard, the Ising terms corresponding to the red edges in the Hamiltonian (\ref{Hamiltonian}), are transformed into new terms $\sum_{<i,j>\in r} \bar {X_i} \bar{X_j}$ on red Ising qubits. It is in fact a new Ising interaction between two red Ising qubits. As shown in Fig. \ref{A3}-c,  these new Ising interactions are matched to a red triangular lattice with red Ising qubits in vertices.
 
In the same way, we apply the basis transformation to other Ising terms of $X_i X_j$ corresponding to edges with other colors in the initial Hamiltonian. It is simple to check that Ising terms corresponding to blue and green edges are also transformed to new Ising interactions between the green and blue Ising qubits living in green and blue plaquettes, respectively. This interactions are matched to green and blue triangular lattices as shown in Fig. \ref{A3}-d. In particular, we notice that there is no interaction between Ising qubits with different colors and we have three decoupled triangular Ising model:
\begin{equation}
H_{\text{II}} = -J_r \sum_{\langle i,j \rangle \in r} \bar{X_i} \bar{X_j} 
               -J_g \sum_{\langle i,j \rangle \in g} \bar{X_i} \bar{X_j} 
               -J_b \sum_{\langle i,j \rangle \in b} \bar{X_i} \bar{X_j}
\end{equation}
Where $r$, $g$, and $b$ denote the sets of red, green, and blue Ising qubits, respectively. Each term represents an Ising interaction on a triangular lattice formed by Ising qubits of the corresponding color.

Finally, by rewriting all terms in the initial Hamiltonian (\ref{Hamiltonian}) in the Ising basis, we have:
\begin{align}
H =\; & H_0 + H_I + H_{II} = \notag\\
     &-N -   ({J_r}\sum_{\langle i,j \rangle \in r} \bar{X}_i \bar{X}_j + ({1- J_r})\sum_{i \in r} \bar{Z}_i) \notag\\
  &- ({J_g}\sum_{\langle i,j \rangle \in g} \bar{X}_i \bar{X}_j + ({1- J_g})\sum_{i \in g} \bar{Z}_i)  \notag\\
     &- ( {J_b}\sum_{\langle i,j \rangle \in b} \bar{X}_i \bar{X}_j + ({1- J_b})\sum_{i \in b} \bar{Z}_i) \label{hamiltonian}
\end{align}

Here $\bar{Z_i}$ and  $\bar{X_i}$ denote logical Pauli operators acting on the Ising qubit associated with plaquette $i$. This Hamiltonian includes three decoupled Ising models in transverse field which are defined on three triangular lattices.

\begin{figure}[h!]
 	\centering
 	\includegraphics[width=8cm,height=5cm,angle=0]{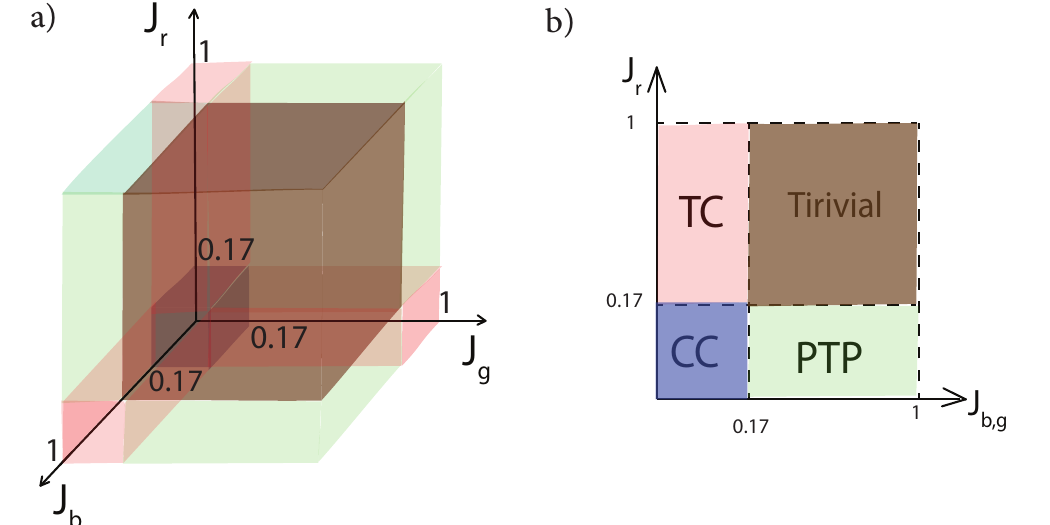}
 	\caption{(Color online) a) A three dimensional phase space for the perturbed Color Code. Regarding mapping to the transverse-field Ising models, planes of $J_r =0.17$, $J_b =0.17$ and $J_g =0.17$ separate eight regions corresponding to different quantum phases. However, regarding symmetries, regions shown with the same color are in the same quantum phase. b) there are four distinct phases in the model. It can be better shown in two-dimensional phase space. CC refers to the Color Code phase, and PTP refers to a partially topological phase.}\label{A4}
 \end{figure}

 It is known that each transverse-field Ising model undergoes a quantum phase transition from a paramagnetic to a ferromagnetic phase at a critical coupling parameter, typically for triangular lattice at $J_c = J^*_c\approx 0.17$ \cite{Blote}. Therefore, we will have a simple three-dimensional phase diagram for the initial model where phase transitions occur across the coordinate planes $J_r = J^*_r$, $J_g =  J^*_g$, and $J_b =  J^*_b$, as shown in Fig. \ref{A4}-a. These planes divide the parameter space into eight regions, and each region represents a physical phase. However, regarding symmetry of model in terms of $J_r$, $J_b$ and $J_g$, there are only four certain phases and other regions show the same phase. In Fig. \ref{A4}-a, we have shown regions which have similar phases with the same colors. In order to better represent four certain phases of the model, it is useful to consider a two-dimemsional phase space by considering $J_b =J_g$. In Fig. \ref{A4}-b, we show such 2D phase diagram. In the following we should characterize nature of each phase in this diagram.

  To characterize the nature of different phases, let us come back to the initial Hamiltonian (\ref{Hamiltonian}). We begin by considering the origin point $(J_r, J_g, J_b) = (0, 0, 0)$ corresponding to a regime in which there is no Ising interaction in the initial Hamiltonian and we have a pure Color Code. Therefore, the first region denoted by "CC" in Fig. \ref{A4}-b represents the Color Code phase. On the other hand, in the transformed Hamiltonian (\ref{hamiltonian}), $J_c$'s are couplings of Ising interactions between Ising qubits and therefore, in $(J_r, J_g, J_b) = (0, 0, 0)$ all three quantum Ising models are in the paramagetic phase. It means that the Color Code phase in the initial Hamiltonian corresponds to three paramagnetic phases in the transformed Hamiltonian.
  
  Next, we consider another region denoted by "Trivial" in Fig. \ref{A4}-b. In particular, when all coupling parameters take the value $(J_r, J_g, J_b) = (1, 1, 1)$, The initial Hamiltonian reduces to the following form:
   \begin{equation}\label{}
   	\begin{aligned}
   		H = - \sum_{p=r,g,b} B_p^{x}  - \sum_{c=r,g,b} \sum_{\langle i,j \rangle \in c} X_i X_j
   	\end{aligned}
   \end{equation}
   The ground state of this Hamiltonian is simply a product state $\prod_i (1+X_i)|0\rangle ^{\otimes N}$ which represent a topologically trivial phase. Notice that $\prod_i (1+X_i)$ is equal to a superposition of all $X$-type strings and therefore, the ground state can be interpreted as a condensed state of strings. In other words, in this phase all $X$-type anyons are condensed. In particular, if we move on the line of $J_r =J_b =J_g$ at the critical point of $0.17$, we observe a phase transition from the Color Code phase to the trivial phase where Ising interactions lead to anyon condensation in the Color Code. We finally notice that in the transformed Hamiltonian (\ref{hamiltonian}), corresponding to the above phase, all three quantum Ising models are in the ferromagnetic phase. In this regard, the trivial phase in the perturbed Color Code corresponds to three ferromagnetic phases in the final transverse-field Ising models.

  It remains to consider two other quantum phases in the phase diagram Fig. \ref{A4}-b. Characterizing these phases needs to more complex analysis that we will do in the next sections.

\begin{figure*}[htbp]
\centering
\includegraphics[width=16cm,height=6cm,angle=0]{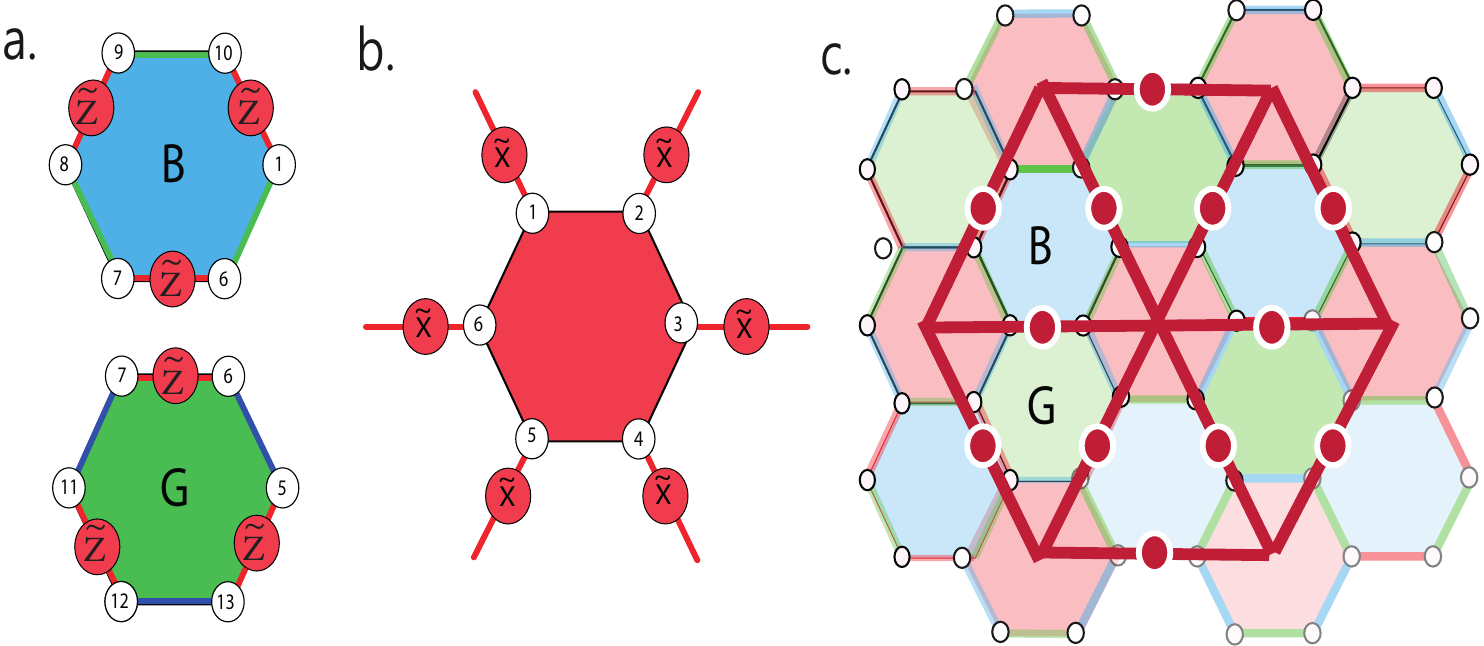}
\caption{(Color online) a) $X$-type stabilizers corresponding to blue and green plaquettes have two qubits in joint with red edges and therefore, they have trivial effect in the green basis. On the other hand, six-body $Z$type stabilizers are mapped to new three-body phase operators in the red basis. b) $X$-type stabilizers are mapped to new six-body shift operators in the red basis. c) Transformed stabilizers in the red basis are matched to stabilizers of a Toric Code model defined on a triangular lattice with red qubits living in edges. }\label{A5}
\end{figure*}
\section{Partial anyon condensation: transition to the Toric Code}\label{sec20} 
 We focus our study in this section on the region denoted by "TC" in Fig. \ref{A4}-b. In particular, consider a vertical path in the 2D phase diagram of Fig. \ref{A4}-b. Starting point is the Color Code phase with $Z_2\times Z_2$ topological order at $(J_r, J_g, J_b) = (0, 0, 0)$. Then, by increasing $J_r$ while keeping $J_{g,b}=0$, the system undergoes a topological phase transition to a new phase at a critical value $ J^*_r$.  We show that this phase is equivalent to the Toric Code, which exhibits $Z_2$ topological order. To characterize this phase, it is enough to consider the initial Hamiltonian for the limiting values of couplings $(J_r, J_g, J_b) = (1, 0, 0)$ where the Hamiltonian (\ref{Hamiltonian}) simplifies to:
\begin{equation}\label{H2}
\begin{aligned}
H = & -\sum_{g,r,b} B_p^{x} 
    - \sum_{b} B_p^{z} - \sum_{g} B_p^{z} 
    - \sum_{\langle i,j \rangle \in r} X_i X_j
\end{aligned}
\end{equation}
First, notice that all terms in the above Hamiltonian commute with each other. Therefore, the ground state of this Hamiltonian is simply written in the following form, up to a normalization factor:
\begin{equation}
	\prod_{p\in g,b}(1+B_p ^x ) \prod_{\langle i,j \rangle \in r}(1+X_i X_j)|0\rangle ^{\otimes N}
\end{equation}
We notice that $\prod_{p\in g,b}(1+B_p ^x )$ is a superposition of blue and green loops while $\prod_{\langle i,j \rangle \in r}(1+X_i X_j)$ is a superposition of red loops as well as red open strings. Therefore, the ground state corresponds to a condensed state of red anyons $r_x$. It means that increasing $J_r$ in the initial Hamiltonian leads to an anyon condensation corresponds to $r_x$'s. Now, we show that this state is in fact equivalent with the Toric Code state. To this end, we show the Hamiltonian (\ref{H2}) is equivalent to the Toric Code Hamiltonian when we apply a change of basis in the following form:
\begin{equation}\label{B2}
   \prod_{\langle i,j\rangle \in r} | \tilde l_{ i,j} \ra =\prod_{\langle i,j\rangle \in r}
    \left( 1 + (-1)^{\tilde  l_{ i,j}} Z_i Z_{j} \right) |{+}\ra^{\otimes N} ,
\end{equation}

 where $\langle i,j\rangle \in r$ referes to qubits living in endpoints of red edges and $\tilde l_{ i,j }$ refers to a binary variable attached to each red link in the hexagonal lattice, see Fig. \ref{A5}. We should rewrite each term of the Hamiltonian (\ref{H2}) in the basis (\ref{B2}). For clarity, we call this set of basis states by the red basis::

1. First notice that in the (\ref{H2}), we have Ising terms of $X_i X_{j}$ corresponding to red edges. Since these operators commutes with $Z_i Z_{j}$ in definition of the the red basis (\ref{B2}), it is concluded that $X_i X_{j} | \tilde l_{i,j}\ra=(+1)|\tilde l_{i,j}\ra$. It means that Ising terms in the new basis are equal to identity operators.

2. Next, we consider our transformation on operators $B_p^x$ corresponding to blue and green Plaquettes. For example, consider a blue plaquette labeled by $B$ in Fig. \ref{A5}-a, where $B_p^x=(X_{10} X_1)(X_6 X_7)(X_8 X_9)$. Since a blue plaquette have two qubit in joint with each red edge, $B_p^x$ commute with $Z_i Z_j$ corresponding to red edges in the red basis. Therefore, This operator is also equal to identity operator in the red basis. A similar result holds for the $B_p^x$ on green plaquettes. For example, $B_p^x=(X_7 X_6)(X_5 X_{13})(X_{12} X_{11})$ corresponding to a green plaquette labeled by $G$ in Fig. \ref{A5}-a has also two qubits in joint with each red edge.

3. situation is different for operators $B_p^x$ corresponding to red plaquettes. As shown in Fig. \ref{A5}-b, each qubit of a red plaquette is joint with a red edge. In this regard, an operator $X_i$ in the $B_p^x$ unti-commute with $Z_i Z_j$ corresponding to a red edge in the red basis (\ref{B2}). In this regard, we conclude that $X_i (1+(-1)^{\tilde l_{ i,j}}Z_i Z_j)=(1+(-1)^{\tilde l_{ i,j} +1}Z_i Z_j)X_i$ and therefore, $X_i|\tilde l_{ i,j}\rangle=|\tilde l_{ i,j} +1\rangle$. It means that if we insert a logical red qubit, that we call it a red link qubit, on the red edge corresponding to $\tilde l_{ i,j}$, $X_i$ plays the role of a shift operator on this qubit. Consequently, as shown in Fig. \ref{A5}-b, the red plaquette operator ${B_p^x}=  X_1  X_2  X_3  X_4  X_5  X_6$ is transformed as $\tilde{B_p^x}= \tilde X_1 \tilde X_2 \tilde X_3 \tilde X_4 \tilde X_5 \tilde X_6$ on six red link qubits around the plaquette $p$. In this regard, in the red basis (\ref{B2}), the first term of the Hamiltonian (\ref{H2}) corresponds to six-body $X$-type logical operators acting on the red link qubits which are located on the red edges. 

4. Finally, we consider the second term in the Hamiltonian (\ref{H2}) which consists of $Z$-type plaquette operators acting on blue and green plaquettes. To examine the effect of $B_p^z$ on blue and green plaquettes, we consider again two representative examples labeled $B$ and $G$ in Fig. \ref{A5}-a. For the blue plaquette $B$, we consider the operator $B_p^z = (Z_{10} Z_1)(Z_6 Z_7)(Z_8 Z_9)$. We notice that this plaquette is surrounded by three red edges in the sense that $Z_{10} Z_1$, $Z_6 Z_7$ and $Z_8 Z_9$ correspond to three red edges. In this regard, when we apply $B_p ^z$ on the red basis (\ref{B2}), $Z_{10} Z_1$, $Z_6 Z_7$ and $Z_8 Z_9$ are applied to $(1+(-1)^ {\tilde l_{10,1}} Z_{10}Z_1)$, $(1+(-1)^{\tilde l_{6,7}} Z_{6}Z_7)$ and $(1+(-1)^{\tilde l_{8,9}} Z_{8}Z_9)$, respectively. Then, by the fact that $Z_{i}Z_j(1+(-1)^{\tilde l_{i,j}}  Z_{i}Z_j)=(-1)^{\tilde l_{i,j}} (1+(-1)^{\tilde l_{i,j} } Z_{i}Z_j)$, we conclude that each $Z_{i}Z_j$ plays the role of a phase operator $\tilde{Z}$ on the corresponding red link qubit. In this regard, $B_p ^z$ in the red basis is transformed to  a new logical operator $\tilde{B}_p^z = \tilde{Z}_{10} \tilde{Z}_6 \tilde{Z}_8$, as shown in Fig. \ref{A5}-a. In the same way, for the green plaquette $G$, the operator $B_p^z = (Z_7 Z_6)(Z_5 Z_{13})(Z_{12} Z_{11})$ reduces to a three-body $Z$-type logical operator $\tilde{B}_p^z = \tilde{Z}_7 \tilde{Z}_5 \tilde{Z}_{12}$ acting on the red link qubits located on red edges around that plaquette.

Thus, in the red basis (\ref{B2}), the Hamiltonian (\ref{H2}) consists of two types of three-body $Z$-type operators and a six-body $X$-type operator, forming the structure of the triangular Toric Code, as illustrated in Fig. \ref{A5}-c. It means that the Hamiltonian (\ref{H2}) is equal to a Toric Code model and accordingly we proved that increasing $J_r$ in the perturbed Color Code leads to a phase transition to the Toric Code state. We emphasize that we can repeat the above argument for condensation of $b_x$ or $g_x$ by increasing $J_b$ or $J_g$. In this regard, in the 3D phase space Fig. \ref{A4}-a there are three regions corresponding to the Toric Code phase.

\section{Characterizing a partially topological phase}\label{sec3}

It remains to consider final region in the 2D phase space Fig. \ref{A4}-b denoted by "PTP". In particular, consider a horizontal path where we start from the Color Code phase at $(J_r, J_g, J_b) = (0, 0, 0)$ and increase $J_g$ and $J_b$ while keeping $J_r = 0$. Along this path, the system undergoes a phase transition at a critical point $(J_g^*=J_b^*)$ toward a new quantum phase. This phase should be different with the Toric Code phase because two types of anyons $b_x$ and $g_x$ are condensed. We show that this phase corresponds to a partially topological phase (PTP). In order to characterize nature of this phase we consider the limiting point of $(J_r, J_g, J_b) = (0, 1, 1)$ where the Hamiltonian (\ref{Hamiltonian}) is simplified to the following form:

\begin{equation}\label{H3}
H =-\sum_{g,r,b} B_p^{x}- \sum_{r} B_p^{z} 
-\sum_{<i,j>\in {g}} X_i X_j -\sum_{<i,j>\in {b}} X_i X_j
\end{equation}

\begin{figure*}[htbp]
\centering
\includegraphics[width=16cm,height=8cm,angle=0]{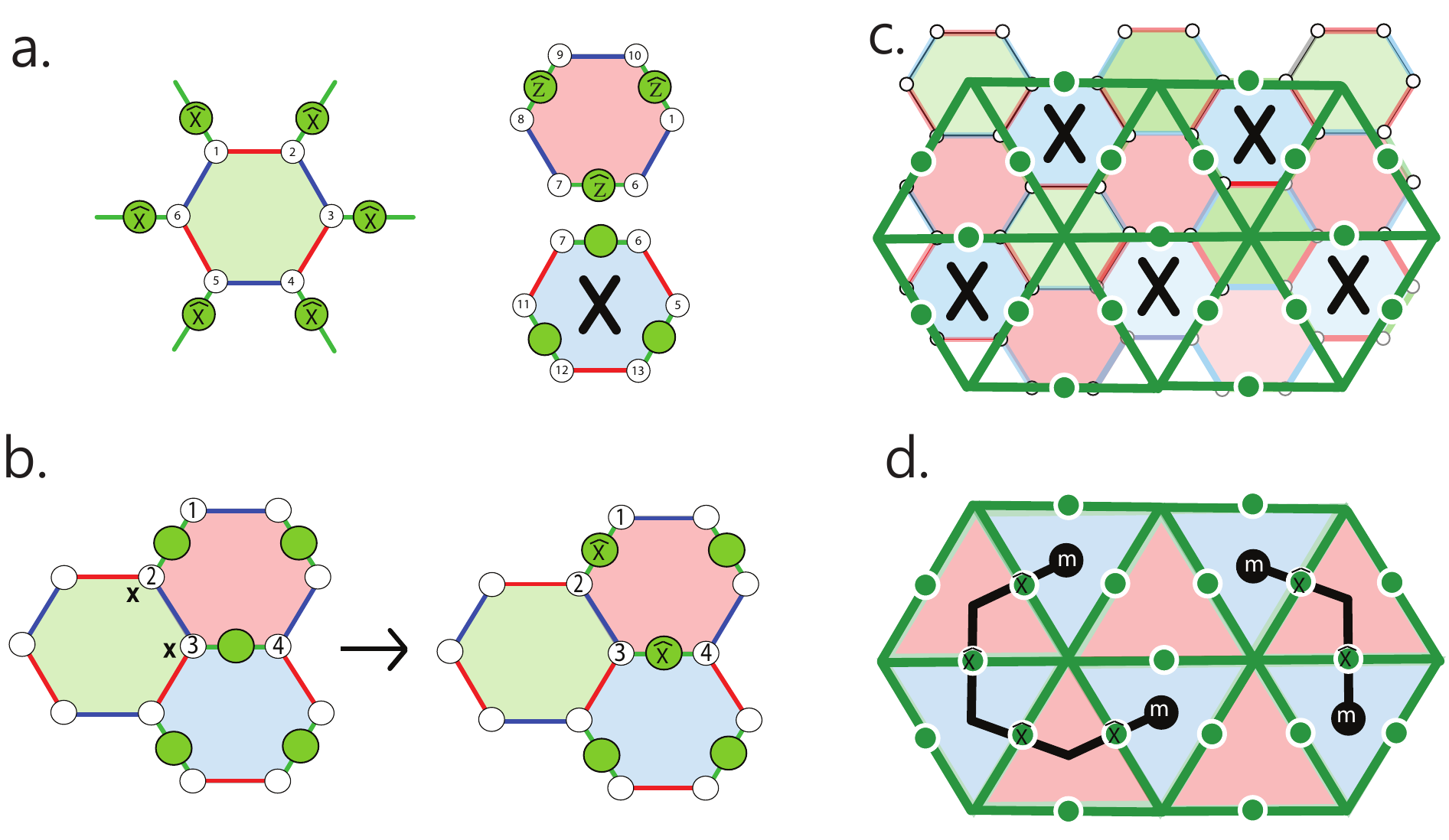}
\caption{(Color online) a) $X$-type stabilizers, corresponding to green plaquettes, are mapped to six-body shift operators in the green basis.  $Z$-type stabilizers corresponding to red plaquettes are mapped to three-body phase operators in the green basis. We notice that there is no $Z$-type plaquette operator in the Hamiltonian corresponding to blue plaquettes. b) An Ising interaction between physical qubits corresponding to blue edges is mapped to an Ising interaction between green qubits belonging to the red plaquettes. c) Transformed stabilizers in the green basis are matched to a perturbed Toric Code on a green triangular lattice. However, there is no plaquette operator corresponding to triangular blue plaquettes. It means that there are holes in the Toric Code corresponding to blue plaquettes. d) New Ising interactions between green qubits can be interpreted as partial condensation of flux anyons in the sense that in the ground state there are open strings whose end-points live only in blue plaquettes. }\label{A6}
\end{figure*}
Notice that the Ising interactions in the above Hamiltonian correspond to the green and blue edges. Similar to the previous section, here we define a new basis to rewrite the above Hamiltonian. To this end, we define green link qubits ${}^{l}_{i,j}$on the green edges, see Fig. \ref{A6}, and consider a basis similar to the (\ref{B2}) with a different that $Z_i Z_j$'s are defined on green edges instead of red edges:
\begin{equation}\label{}
	\prod_{\langle i,j\rangle \in g} |\,{}^{l}_{i,j} \rangle
	\left( 1 + (-1)^{{ l}_{ i,j}} Z_i Z_{j} \right) |{+}\ra^{\otimes N} ,
\end{equation}

 Then, different terms in the Hamiltonian (\ref{H3}) must be rewritten in the new basis. For clarity, we call this set of basis states by the green basis:

1. Since operators $B_p^x$ corresponding to red and blue plaquettes commute with $Z_i Z_j$ corresponding to the green edges, the effect of these terms in the green basis is trivial and they can replaced with identity operators.

2. For $B_p^x$'s corresponding to the green plaquettes, We notice that, as shown in Fig. \ref{A6}-a, they have only one qubits in joint with green edges and therefore, $B_p^x$ unti-commute with $Z_i Z_j$ corresponding to the green edges. In particular, $X_i$ plays the role of a logical shift operator on the corresponding green link qubit. In this regard, the operator $B_p^x= X_1 X_2 X_3 X_4 X_5 X_6$ shown in Fig. \ref{A6}-a, is transformed to $B_p^x = \hat{X}_1 \, \hat{X}_2 \, \hat{X}_3 \, \hat{X}_4 \, \hat{X}_5 \, \hat{X}_6$ defined on six green link qubits on the green edges as shown in Fig. \ref{A6}-a.

3. Next, we consider operators $B_p^z$ corresponding to red Plaquettes. To this end, consider a red plaquette illustrated in Fig. \ref{A6}-a. Notice that a red plaquette has three green edges and therefore, the operator $B_p^z = (Z_{10} Z_1)( Z_6 Z_7)( Z_8 Z_9)$ can be expressed as a product of $Z_i Z_{j}$ terms acting on green edges. Then, since the effect of $Z_i Z_{j}$ on the green basis is equal to a logical phase operator $\hat{Z}_i$, it is concluded that $B_p ^z$ is transformed to a logical three-body operator $\hat{B}_{p}^z =\hat{Z}_{10} \hat{Z}_6 \hat{Z}_8$. We also emphasize that in the initial Hamiltonian (\ref{H3}), we have no operator $B_p^z$ corresponding to blue plaquettes. Therefore, as shown in Fig. \ref{A6}-a, there is no three-body $Z$-type term in the transformed Hamiltonian on green qubits around a blue plaquette.

 4. Finally, we consider Ising terms $X_i X_j$ corresponding to green edges. Since they commute with $Z_i Z_j$ defined in the green basis, it is clear that their effect on the green basis is trivial. On the other hand, since each blue edge shares only one physical qubit with an adjacent green edge, $X_i X_j$ terms corresponding to blue edges would have a non trivial effect on the green basis. As shown in Fig. \ref{A6}-b, consider a particular blue edge with Ising term of $X_2 X_3$. This operator unti-commutes with operators $Z_1 Z_2$ and $Z_3 Z_4$ corresponding to adjacent green edges. Therefore the effect of $X_2 X_3$ in the green basis would be equal to  the logical operator $\hat{X}_{1} \hat{X}_{3}$ which is applied to two green link qubits as shown in Fig. \ref{A6}-b.
 
 Regarding the above transformations, the Hamiltonian (\ref{H3}) is mapped to the following Hamiltonian defined on green link qubits, up to a constant term:
 
 \begin{equation}\label{H4}
 	\hat{H}=-\sum_{p\in G}\hat{B}_{p}^x -\sum_{p \in R}\hat{B}_{P}^z -\sum_{\langle i,j\rangle \in G}\hat{X}_i \hat{X}_j
 \end{equation}
 
 Two the first terms are well matched to a triangular Toric Code as shown in Fig. \ref{A6}-c where $\hat{B}_{p}^x$ and $\hat{B}_{P}^z$ correspond to vertex and plaquette operators of the Toric Code, respectively. However, we notice that there is no $B_p ^z$ operator corresponding to the blue plaquettes of the hexagonal lattice in the (\ref{H3}). Therefore, in the final triangular Toric Code model we do not have three-body plaquette operators corresponding to triangular plaquettes which are marked with a cross in the Fig. \ref{A6}-b. It means that we have a Toric Code model with holes which are designed in a checkerboard pattern, see Fig. \ref{A6}-c. 
 
 The above interesting point is better clarified when we notice to the last terms in the transformed Hamiltonian (\ref{H4}). Since all terms in the (\ref{H4}) commute with each other, the ground state of this Hamiltonian is written in the following form:
 \begin{equation}\label{goo}
 \prod_{P \in G}(1+\hat{B}_{P}^x)\prod_{\langle i,j\rangle \in G}(1+\hat{X}_i \hat{X}_j ) |\hat{0} \hat{0} ...\hat{0}\rangle ,
 \end{equation}
  where $\langle i,j\rangle \in G$ refers to two green qubits belonging to a red triangle in Fig. \ref{A6}-d where final Toric Code is defined. In this regard, notice that $\prod_{\langle i,j\rangle \in G}(1+\hat{X}_i \hat{X}_j )$ is equal to a superposition of $X$-type string operators, whose end-points live in blue plaquettes. In fact, each $\hat{X}_i \hat{X}_j$ generates two $m$-type anyons at two blue plaquettes as shown in Fig. \ref{A6}-d. In this regard, each porduct of $\hat{X}_i \hat{X}_j$ in the phrase of $\prod_{\langle i,j\rangle \in G}(1+\hat{X}_i \hat{X}_j )$, also generates $m$-type anyons which move in blue plaquettes. In other words, the ground state of the model contains $m$-type anyons that are partially confined to move in the blue plaquettes. It is interesting to describe this phenomena by categorizing the $m$-type anyons in the Toric Code into two distinct subsets: $m_b$ and $m_r$. The $m_b$ anyons live in blue plaquettes while $m_r$'s live in red plaquettes. Then, in the ground state (\ref{goo}), only $m_b$ anyons are condensed while $m_r$ are not free to generate and move in the lattice. Therefore, we call this phenomena a partial anyon condensation in the Toric Code.

In this regard, we identified four different quantum phases in our perturbed Color Code model and used mappings to characterize the nature of different phases. However, it is known that topological nature of topological quantum phases is characterized by non-local order parameters. In the following section, we define the string order parameters and show how they can be employed to probe the topological nature of different phases.

\section{String Order Parameter}\label{sec4}

In this section, we define the string order parameters for the perturbed Color Code model \cite{Zarei3} and investigate how they can be used to distinguish different phases and uncover the underlying topological order. To this end, we use the mapping between perturbed Color Code and transverse-field Ising models presented in Sec. \ref{sec2}. In particular, we know that in the transverse-field Ising model, the average magnetization is a local order parameter that separates the ferromagnetic phase from the paramagnetic one. In this regard, here we show that there are string order parameters in the perturbed Color Code model which are mapped to Ising order parameters in transverse-field Ising models. We use such mapping to analyze the behavior of string order parameters for different phases of our model. 
\begin{figure}[h!]
\centering
\includegraphics[width=8cm,height=4cm,angle=0]{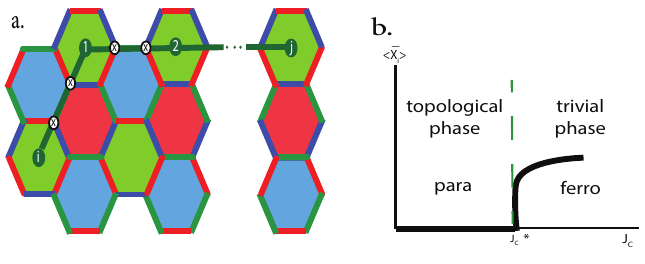}
\caption{ (Color Online) a) Visualization of the green string order parameters in the Color Code. A correlation function on Ising qubits $\la \bar{X}_i \bar{X}_j \ra$ is mapped to a string operator on physical qubits in the perturbed Color Code.  b) A schematic phase diagram of the transverse-field Ising model which shows a phase transition at $J_c=J_c^*$. The paramagnetic (ferromagnetic) phase in the dual picture corresponds to the topological (trivial) phase in the original model.} \label{A7}
\end{figure} 

To establish the above mapping, we start from three transverse-field Ising models defined on three triangular lattices as presented in Sec. \ref{sec2}. We remind that Ising qubits live in center of hexagonal plaquettes of the Color Code. Now, as shown in Fig. \ref{A7}-a, consider two plaquettes $i$ and $j$ with the same color, for example green color in Fig. \ref{A7}-a. There are two green Ising qubits in the above plaquettes and we define an operator $  \bar X_i \bar X_j $ applied in the above qubits. Notice that in transverse-field Ising model defined in the green triangular lattice, the expectation value $ \langle \bar X_i \bar X_j \rangle $ is a correlation function. However, at large distances between $i$ and $j$, it is known that $ \langle \bar X_i \bar X_j \rangle= \langle \bar X_i \rangle \langle \bar X_j \rangle $. On the other hand, $\langle \bar X_i \rangle$ is the same as magnetization $M$ in the transverse-field Ising model and therefore it is concluded that $\langle \bar X_i \bar X_j \rangle= M^2$.

Next, we notice that the operator $\bar X_i \bar X_j$ can be written as a product of nearest-neighbor Ising terms along a path connecting $i$ to $j$ in the following form, as shown in Fig. \ref{A7}-a:
 \begin{equation}\label{po}
 	\bar X_i \bar X_j = (\bar X_i \bar X_1) (\bar X_1 \bar X_2) .... (\bar X_N \bar X_j)
 \end{equation} 
  
 Notice that each $\bar X_k \bar X_{k+1}$  is a Ising interaction between Ising qubits located on neighboring plaquettes of the same color $c$. On the other hand, we remind that the operator of $\bar X_k \bar X_{k+1}$ on Ising qubits is equal to an operator $X_k X_{k+1}$ on physical qubits of the Color Code. In other words, when $X_k X_{k+1}$ on physical qubits, which live in two end-points of an green edge, is rewritten in the Ising basis (\ref{basis1}), it is transformed to $\bar X_k \bar X_{k+1}$ on Ising qubits which live in the corresponding green plaquettes, see Fig. \ref{A3}-b. In this regard, by using such transformation for all Ising terms in (\ref{po}), the operator $\bar X_i \bar X_j$ maps back to a product of Pauli-X operators along a string of edges with color $c$ in the perturbed Color Code model in the form of $S_c=\prod_{i\in c}X_i$ where $X$'s are applied on physical qubits. Thus, each Ising magnetization $m_c$ corresponding to transverse-field Ising model on triangular lattice with color $c$, is mapped to a string order parameter $S_c$ corresponding to string with color $c$ in the initial model.
   
 On the other hand, we remind that paramagnetic phase in the Ising model corresponds to topological phase in the Color Code model. In particular, in the paramagnetic phase $m_c =0$ and therefore in the topological phase we have $S_c =0$. In other words, zero value for $S_c$ is a signature of topological order while in the trivial phase $S_c$ has a non-zero value corresponding to ferromagnetic phase in Ising model. In summary, ferromagnetic (ordered) and paramagnetic (disordered) phases in the Ising picture are mapped respectively to topological and trivial phases in the original model, see Fig. \ref{A7}-b. However, notice that there are three string order parameters in the Color Code model and by tuning the coupling constants $J_r$, $J_g$, and $J_b$, the system exhibits a rich phase structure characterized by the presence or absence of topological order as summarized in Table~\ref{tab:phase-diagram}.  When all couplings are maximal ($J_r=J_b=J_g=1$ ),  all three transverse-field Ising models in the dual picture are in the ferromagnetic phase. Consequently, all three string order parameters are equal to 1, indicating that the system is in the trivial phase. On the other hand, when all three couplings vanish, i.e., $J_r=J_b=J_g=0$, the transverse-field Ising models are in paramagnetic phases, and all string order parameters drop to zero. In this limit, the system enters the topological Color Code phase.
\begin{table}[h!]
\centering
\begin{tabular}{|c|c|c|c|c|c|c|c|}
\hline
$J_r$ & $J_g$ & $J_b$ & Ising Phases & Topological Phases & $S_r$ & $S_g$ & $S_b$ \\
\hline
1 & 1 & 1 & F (r,g,b) & Trivial & 1 & 1 & 1 \\
0 & 0 & 0 & P (r,g,b) & Topological Color Code & $0$ & $0$ & $0$ \\
1 & 0 & 0 & F (r), P (g,b) & Toric Code (red) & $1$ & 0 & 0 \\
0 & 1 & 0 & F (g), P (r,b) & Toric Code (green) & 0 & $1$ & 0 \\
0 & 0 & 1 & F (b), P (r,g) & Toric Code (blue) & 0 & 0 & $1$ \\
1 & 1 & 0 & F (r,g), P (b) & Partially Topological & $1$ & $1$ & 0 \\
1 & 0 & 1 & F (r,b), P (g) & Partially Topological & $1$ & 0 & $1$ \\
0 & 1 & 1 & F (g,b), P (r) & Partially Topological & 0 & $1$ & $1$ \\
\hline
\end{tabular}
\caption{Phase classification based on the values of $J_r$, $J_g$, and $J_b$.}
\label{tab:phase-diagram}
\end{table}

In the case where only one of the couplings is non-zero— for example, ($J_r=1, J_b=J_g=0$)—only the red Ising model is in the ferromagnetic phase, while the other two models remain in paramagnetic phase. As we explained in Sec. \ref{sec20}, these phase corresponds to the Toric Code model defined on red qubits. Therefore, in the Toric Code phase, one of string operators $S_r$ has non-zero value while $S_b =S_g =0$. In other words, in the initial Color Code, non-zero value of $S_r$ shows that red anyons are condensed and the system shows a phase transition to the Toric Code phase. Similarly, there are also two other Toric Code phases corresponding to condensation of blue or green anyons. 

On the other hand, when two coupling parameters are non-zero and one of them is zero e.g., $J_r=0, J_b=J_g=1$, two of the Ising models are in the ferromagnetic phases while one remains in the paramagnetic phase. This leads to a different phase in which two of the string order parameters are non-zero and one is zero. In other words, anyons corresponding to two of colors are condensed and we have a partially topological phase as we explained in Sec. \ref{sec3}. In particular, notice that in this phase one of the string order parameters is equal to zero. Since zero value for string order parameters was a signature of the topological order, we conclude that this phase has still a type of topological order. It is our reason that we called this phase a partially topological phase.

\section{Discussion}
We introduced a Hamiltonian approach in which color-dependent Ising perturbations, added to the Color Code Hamiltonian, lead to various topological phase transitions. Through a basis transformation, we showed that the model is mapped onto three decoupled transverse-field Ising models corresponding to three colors. This mapping not only allowed us to understand the phase transitions using standard Ising critical behavior, but also it revealed a rich structure of topological phases. 
In particular, by analyzing the string order parameters, we characterized the emergence and breakdown of topological order as Ising interactions are varied. In this regard, we identified both fully topological and trivial phases, and also uncovered a partially topological phase which exhibits a weaker form of topological order compared to the Color Code and the Toric Code. As a concluding remark, we would like to emphasis that partial anyon condensation represents a conceptual extension of conventional condensation frameworks, which typically consider full reduction to known topological phases. Here, we have shown that controlled perturbations can generate partially topological phases with reduced topological properties. In this regard, our study highlights the power of a Hamiltonian-based viewpoint in uncovering novel topological phenomena.

\end{document}